\documentclass[aps,pra,twocolumn,showpacs,superscriptaddress,groupedaddress]{revtex4}
\usepackage{amsmath}
\usepackage{graphicx}
\usepackage{dcolumn}
\usepackage{bm}
\usepackage{amssymb}

\begin{document}

\title{Newton's cradles in optics: From to $N$-soliton fission to soliton
chains}
\author{R. Driben}
\email{driben@post.tau.ac.il}
\affiliation{Department of Physical Electronics, School of Electrical Engineering,
Faculty of Engineering, Tel Aviv University, Tel Aviv 69978, Israel}
\affiliation{Department of Physics CeOPP, University of Paderborn, Warburger Str. 100,
D-33098 Paderborn, Germany}
\author{B. A. Malomed}
\affiliation{Department of Physical Electronics, School of Electrical Engineering,
Faculty of Engineering, Tel Aviv University, Tel Aviv 69978, Israel}
\author{A. V. Yulin}
\affiliation{Centro de Fisica Teorica e Computacional, Faculdade de Ciencias,
Universidade de Lisboa, Avenida Professor Gama Pinto 2, Lisboa 1649-003,
Portugal}
\author{D. V. Skryabin}
\affiliation{Department of Physics, University of Bath, Bath BA2 7AY, United Kingdom}
\date{\today}

\begin{abstract}
A mechanism for creating a Newton's cradle (NC) in nonlinear light
wavetrains under the action of the third-order dispersion (TOD) is
demonstrated. The formation of the NC structure plays an important
role in the process of fission of higher-order ($N$-) solitons in
optical fibers. After the splitting of the initial $N$-soliton into
a nonuniform chain of fundamental quasi-solitons, the tallest one
travels along the entire chain, through consecutive collisions with
other solitons, and then escapes, while the remaining chain of
pulses stays as a bound state, due to the radiation-mediated
interaction between them. Increasing the initial soliton's order,
$N$, leads to the transmission through, and release of additional
solitons with enhanced power, along with the emission of radiation,
which may demonstrate a broadband supercontinuum spectrum. The NC
dynamical regime remains robust in the presence of extra
perturbations, such as the Raman and self-steepening effects, and
dispersions terms above the third order. It is demonstrated that
essentially the same NC mechanism is induced by the TOD in finite
segments of periodic wavetrains (in particular, soliton chains). A
strong difference from the mechanical NC is that the TOD-driven
pulse passing through the soliton array collects energy and momentum
from other solitons. Thus, uniform and non-uniforms arrays of
nonlinear wave pulses offer an essential extension of the mechanical
NC, in which the quasi-particles, unlike mechanical beads, interact
inelastically, exchanging energy and generating radiation.
Nevertheless, the characteristic phenomenology of NC chains may be
clearly identified in these nonlinear-wave settings too.
\end{abstract}

\maketitle

\widetext





\section{Introduction}

Beyond its natural manifestations in various mechanical systems \cite{cradle}%
, the commonly known Newton-cradle (NC) setup (which was actually first
built, as an experimental device, by French physicist Edme Mariotte in 1670
\cite{teacher}) has found microscopic realizations in atomic and molecular
chains \cite{NC}. Another physically interesting possibility is to build NCs
as a chain of solitons or wave pulses, if they may be considered as
repulsively interacting quasiparticles. This option was elaborated in detail
in terms of an immiscible binary Bose-Einstein condensate, assuming that the
chain is built as a string of alternating solitons composed of mutually
repelling components, while the intrinsic nonlinearity of each component is
self-attractive \cite{Ourense}. Still earlier, a similar system was studied
experimentally and theoretically in the form of \textquotedblleft
supersolitons" in a chain of magnetic-flux quanta (fluxons) pinned to a
lattice of defects in a long Josephson junction \cite{super}. Very recently,
it was reported that a finite chain of \textit{dissipative solitons} in a
two-dimensional model of the laser cavity, with a built-in spatially
periodic grating, may also feature an NC dynamical regime, if the trapped
chain is initially hit by a moving soliton \cite{Besse}. Characteristic
features of the NC phenomenology are clearly observed in that setting, even
if the underlying dynamics is dissipative, being governed by a complex
Ginzburg-Landau equation.

In the present work we aim to present a realization of the NC in nonlinear
optical fibers, where the \textquotedblleft cradle", in the form of a
non-uniform train of quasi-solitons, is naturally produced as a result of
the fission of $N$-solitons under the action of the third-order dispersion
(TOD). The process of higher-order soliton fission is, by itself, of great
significance to applications (see below). It will be demonstrated that the
realization of the NC array in this setting is vastly different from the
classical chain of identical hard beads: the constituent solitons have
different sizes, and they interact inelastically, exchanging energy and
emitting considerable amounts of radiation. Nevertheless, the overall
dynamical scenario may be identified as that characteristic to NC systems:
the uttermost soliton starts the motion from the left edge of the array,
which triggers a localized wave of the energy and momentum transfer across
the array, ending up with the release of one or several free solitons from
the right edge, while the other solitons stay in the form of a bound chain.
In fact, these results suggest that the NC concept is not restricted to the
mechanical realizations, and may be extended to more general settings. In
fact, an essential extension suggested by the present results is that
colliding quasi-particles may effectively \emph{pass through each other} in
the NCs based on soliton arrays.

It is well known that, in addition to the ubiquitous fundamental solitons,
the integrable nonlinear Schr\"{o}dinger (NLS) equation and physical systems
described by nearly-integrable versions of this equation \cite%
{KivsharMalomed} support $N$-soliton states, with $N\geq 2$, which are
oscillating modes periodically restoring their shape at distances that are
multiples of the fundamental soliton period \cite{Satsuma}. Although they
are subject to a weak splitting instability, due to the fact that their
binding energy is equal to zero, in the case of the integrable equation (the
splitting may be catalyzed by a weak resonant periodic modulation of the
nonlinearity strength \cite{HS}), higher-order solitons were observed in
nonlinear optical fibers \cite{Observations}, and robust $2$-solitons have
been created as basic modes in soliton lasers \cite{laser}. Recently,
dynamics of higher-order solitons was studied theoretically in regular and
parity-time-symmetric nonlinear dual-core couplers \cite{EPL2012}.

The fission of $N$-solitons in nonlinear optical fibers is naturally caused
by higher-order linear and nonlinear effects \cite{Golovchenko},\cite{Kodama}%
,\cite{Herrmann},\cite{fissionchina}, such as the Raman-induced
self-frequency shift \cite{Mitschke},\cite{Gordon}, self-steepening \cite%
{Blow}, and third-order dispersion (TOD), which are included into the
corresponding generalized NLS equations \cite{Agrawal},\cite{Dudleyobzor}.
The point at which the fission occurs is, usually, the one where the
bandwidth of the evolving $N$-soliton attains its maximum. On the other
hand, the oscillations of the $N$-soliton in its unperturbed form can be
naturally used for strong compression of the pulse (the so-called
soliton-compression effect) \cite{Dianov},\cite{Chen}. The fission of
higher-order solitons is a key enabling mechanism for the generation of
ultrashort frequency-tuned fundamental solitons \cite{Tunning}, and for the
creation of ultra-broadband supercontinuum \cite{Dudleyobzor,SC}. Recent
advances in manufacturing photonic-crystal fibers filled with Raman-inactive
gases \cite{Travers} provide an additional motivation to focus the studies
on the TOD-induced fission and its potential applications to photonics, in
the absence of the Raman effect.

In earlier works, it was demonstrated that higher-order terms, treated as
small perturbations, split $N$-solitons into a set of $N$ different
fundamental solitons, with the sequence of peak powers analytically
predicted by the well-known exact solution of Satsuma and Yajima, see Eq. (%
\ref{SY}) below \cite{Satsuma}, \cite{Kodama}. However, we demonstrate below
that, when the input pulse is injected close to the zero-dispersion point,
hence the TOD effect is relatively strong, such a simple picture of the
soliton fission is far from reality and a detailed numerical study is required.

The rest of the paper is structured as follows. The model is formulated in
Section II, which is followed by the presentation of the main numerical
results in Section III. As mentioned above, the results demonstrate that the
fission of $N$-solitons under the action of the TOD proceeds via the
creation of NCs in the form of nonuniform soliton chains, and passage of one
or several tallest solitons across the entire chain, ending by release of the passing solitons, while the remaining ones stay in
the form of bound array. An essential difference of the NC dynamical regime
in the soliton arrays from the NC in chains of mechanical particles is that
collisions of solitons are partly inelastic, in the nonintegrable system. As
a result, the passing soliton collects additional energy and momentum from
the other pulses, and is eventually released from the array with a
considerable increase of the energy, and with a large frequency shift. To
demonstrate the relevance of the generalized NC concept in the broader
context of nonlinear optics, in Section IV we demonstrate the realization of
the TOD-driven NC in periodic patterns of wave pulses, such as
\textquotedblleft prefabricated" chains of identical solitons. The paper is
concluded by Section IV.

\section{The model}

The dynamics of ultrashort pulses in nonlinear fibers is governed by the
generalized NLS equation for amplitude $A$ of the electromagnetic field,
which includes the TOD, self-steepening, and Raman terms \cite%
{Agrawal,Dudleyobzor}:
\begin{gather}
\frac{\partial A}{\partial z}=\sum\limits_{m\geq 2}\frac{i^{m+1}\beta _{m}}{%
m!}\frac{\partial ^{m}A}{\partial T^{m}}  \notag \\
+i\gamma \left( 1+\frac{i}{\omega _{0}}\frac{\partial }{\partial T}\right) %
\left[ A(z,\tau )\int_{-\infty }^{T}d\tau ^{\prime }R(T-T^{\prime
})|A(z,T^{\prime })|^{2}\right] ,  \label{NLS}
\end{gather}%
where, as usual, $z$ and and $T$ are the propagation distance and reduced
time \cite{Agrawal}, and $\beta _{m}$ is the $m$th-order dispersion
coefficient at carrier frequency $\omega _{0}$. The fiber loss is neglected
in Eq. (\ref{NLS}), as the fission of the $N$-soliton and formation of the
NC occur on the propagation-distance scale which is much shorter than the
absorption length. Further, the nonlinearity coefficient is taken as $\gamma
=0.08$ W$^{-1}$m$^{-1}$, the response function is $R(T)=(1-f_{R})\delta
(T)+f_{R}h_{R}(T)$, with the first and second terms standing for the
instantaneous and delayed (Raman) contributions, respectively. Here $%
f_{R}=0.18$ is the fraction of the Raman contribution to the nonlinear
polarization, and $h_{R}(\tau )=(\tau _{1}^{2}+\tau _{2}^{2})/(\tau _{1}\tau
_{2}^{2})\exp (-T/\tau _{2})\sin (T/\tau _{1})$ approximates the Raman
response function of the silica fiber \cite{Blow}, with $\tau _{1}=12.2$ fs
and $\tau _{2}=32$ fs. Hereafter, the second-order group-velocity-dispersion
(GVD) coefficient is taken as $\beta _{2}=-0.0021$ ps$^{2}$/m, with the
pulse injected at carrier wavelength $\lambda =800$ nm into the
anomalous-GVD region of the fiber, close to the zero-dispersion point (ZDP).

Equation (\ref{NLS}) was solved numerically by means of the split-step
Fourier method \cite{Agrawal}. Actually, the simulations of the $N$-soliton
fission included the Raman effect and the shock term only at the last stage
of the analysis, the results of which are displayed below in Figs. \ref{fig6}
and \ref{fig7} to show that these terms are not crucially important for the
NC\ dynamics, unlike the TOD. Higher-order-dispersion terms, up to the
seventh order, are also included in simulations presented in Fig. \ref{fig7}%
. As mentioned above, neglecting the Raman self-frequency shift is relevant,
in particular, for the pulse propagation in holey fibers filled by
Raman-inactive gases \cite{Travers}, where the effective Raman coefficient
is 4 or 5 orders of magnitude smaller than in silica.

\section{$N$-soliton fission via the Newton-cradle mechanism}

Simulations of Eq. (\ref{NLS}) were performed with the input in the form of
the usual $N$-soliton,%
\begin{equation}
u(z=0,T)=N\sqrt{P_{0}}\mathrm{sech}(T/T_{0}),  \label{N}
\end{equation}%
with $N>1$, and by setting $T_{0}=50$ fs (the respective
FWHM width is $\approx 90$ fs). With the above-mentioned value of the
second-order GVD coefficient $\beta _{2}$ $=-0.0021$ ps$^{2}$/m, the peak
power of the fundamental soliton corresponding to this width is $P_{0}=10.5$
W. While the second-order dispersion parameter is kept constant, its third-order counterpart (the TOD
coefficient) will be varied.
In the case of weak TOD, the analysis of the $N$-soliton splitting, based on
the perturbation theory, was developed in Ref. \cite{Kodama}. However, for
larger values of $\beta _{3}$, the dynamics is more complex, strongly
deviating from the prediction of the perturbative analysis (see Fig. \ref%
{fig4} below). The TOD distorts the fundamental pulse, making it asymmetric,
with a periodic oscillatory structure emerging near one edge \cite{Agrawal, Resonantradiation},
which results from dispersive waves and is described by the Airy function
\cite{Besieris}. Accordingly, the shape of the soliton experiences a strong
deformation with the appearance of this oscillating structure at the front
or rear side of the soliton \cite{Agrawal}, depending on the sign of $\beta
_{3}$. Here we consider the case of opposite signs of the TOD and
second-order GVD (for identical signs, the results can be obtained by
substitution $T\rightarrow -T$, which reverses the sign of $\beta _{3}$).

\begin{figure}[tbp]
\centerline{\includegraphics[scale=0.3]{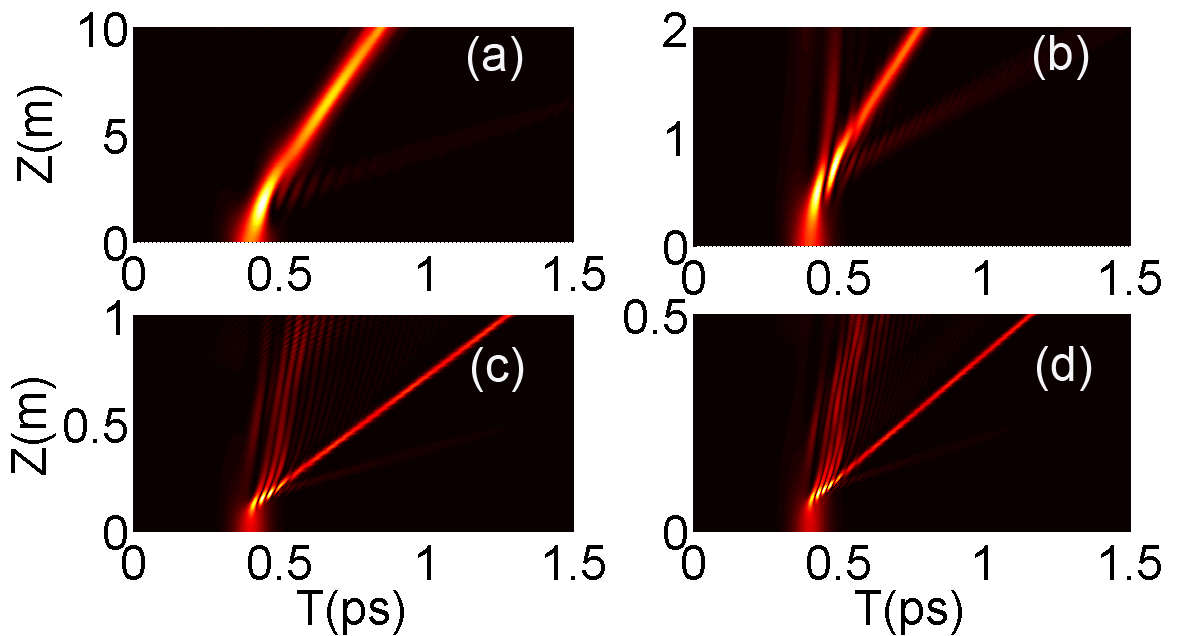}}
\caption{(Color online) The fission dynamics of the $1.5$-soliton (a), $3$%
-soliton (b), $5$-soliton (c), and $10$-soliton (d). The simulations of Eq. (%
\protect\ref{NLS}) with $\protect\beta _{3}=6.98\cdot 10^{-5}$ ps$^{3}$/m
were performed for input (\protect\ref{N}) with $T_{0}=50$ fs and $%
P_{0}=10.55$ W.}
\label{fig1}
\end{figure}

Despite the fact that the energy coupled into the radiation under the action
of TOD is relatively small, it was demonstrated that the dispersive wave
carries away a significant amount of momentum \cite{Wai}. The periodic light structure at the edge of the pulse induces effective refractive index change, that attracts the main peak of the pulse. If more energy is
injected into the fiber than it is required for the formation of the
fundamental soliton, the interaction between the oscillatory structure and
the peak of the pulse becomes stronger, causing acceleration or delay of the
pulse. This situation is shown in Fig. \ref{fig1}(a), where input pulse (\ref%
{N}) was taken with peak power $23.625$ W, corresponding to $N=1.5$ at $%
T_{0}=50$ fs.

The injection of larger energy leads to further enhancement of the
interaction between the main peak and the oscillatory structure. The
first-born tallest soliton starts to advance from the left edge of the
wavetrain to the right under the action of the strong TOD, consecutively
colliding with all the other peaks in the emerging chain. Phase shifts $%
\Delta \varphi $ between adjacent peaks were measured to range in the
interval of $\pi /2<\Delta \varphi <3\pi /2,$ making the collisions between
individual pulses repulsive \cite{KivsharMalomed}. The colliding pulses
exchange the energy and momentum before the soliton is ejected from the
pulse array, carrying extra energy and featuring a frequency shift (i.e.,
carrying extra momentum), which were accumulated due to non-elasticity of
collisions under the action of TOD \cite{Energy}.

\begin{figure}[tbp]
\includegraphics[scale=0.2]{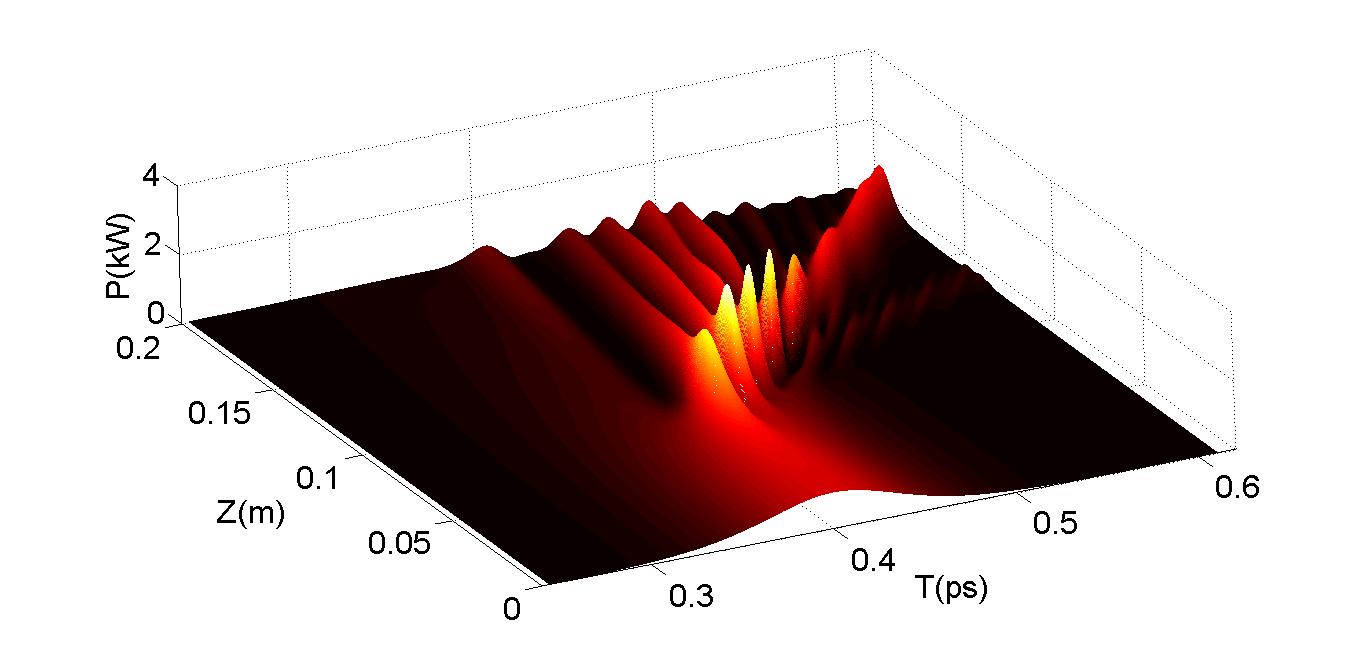} %
\includegraphics[scale=0.4]{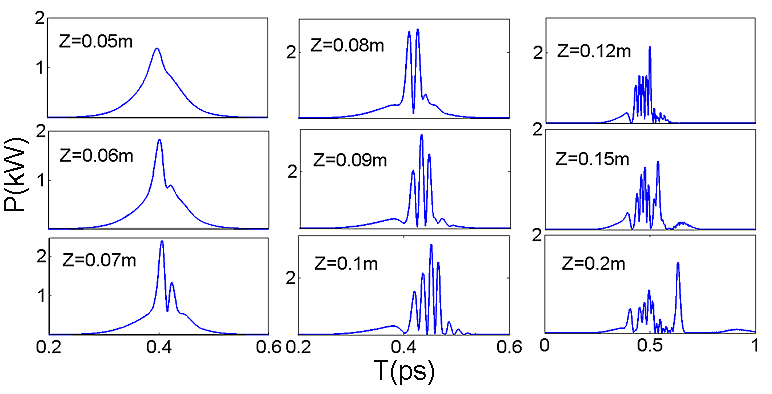}
\caption{(Color online) (a) The zoomed 3D view of the fission of the $10$%
-soliton from Fig. \protect\ref{fig1}(d), in the area where the
\textquotedblleft cradle" emerges. (b) Snapshots of the evolution of
the local-power distribution, $P\equiv \left\vert A\left( z,T\right)
\right\vert ^{2}$. The tallest soliton is ejected at $Z\simeq 0.18$
m.} \label{fig2}
\end{figure}

Figures \ref{fig1}(b)-(d) display the dynamical regimes developed from input
(\ref{N}) corresponding to $N=3,5,$ and $10$, with $T_{0}=50$ fs. In the
case of the $3$-soliton input [Fig. \ref{fig1}(b)], the escaping soliton is
actually ejected from the second channel of the arrayed structure. Further,
the fission of the $5$-soliton input [Fig. \ref{fig1}(c)] results in the
ejection from the third channel, and, in the case of $N=10$ [Fig. \ref{fig1}%
(d)], the ejection from the seventh effective channel is observed. In each
case, additional pulses, located to the right of the ejecting channel,
remain very weak at the moment when the soliton escapes. To further
illustrate the NC\ dynamics following the fission of the $10$-soliton, Fig. %
\ref{fig2}(a) zooms the area from Fig. \ref{fig1}(d) where the
\textquotedblleft cradle" emerges, and Fig. \ref{fig2}(b) displays snapshots
of the field's intensity at several values of the
propagation distance, $Z$.

\begin{figure}[tbp]
\includegraphics[scale=0.4]{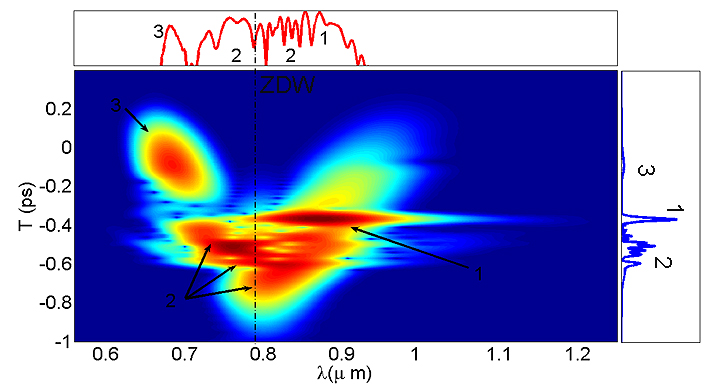} %
\includegraphics[scale=0.4]{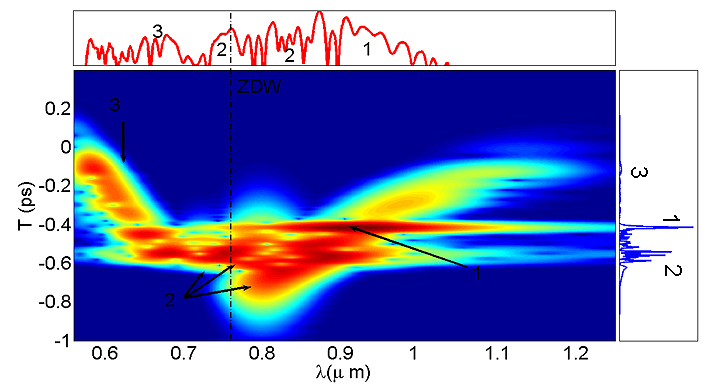}
\caption{(Color online) XFROG representation of the $10$-soliton-fission
dynamics at $Z=0.2$ m with (a) $\protect\beta _{3}=6.98\cdot 10^{-5}$ ps$^{3}
$/m, and (b) $\protect\beta _{3}=1.75\cdot 10^{-5}$ ps$^{3}$/m.}
\label{fig3}
\end{figure}

While the tallest soliton, driven by the TOD, passes the whole chain from
left to right and eventually escapes, the remaining pulses continue the
propagation in a bound state. In the case of large values of the TOD
coefficient, $\beta _{3}$, the chain mostly stays in the region of the
normal second-order GVD, while, with the decrease of $\beta _{3}$, it is
shifted mostly to the anomalous-GVD domain, thus increasing the soliton
content of the chain. The situation is illustrated by Fig. \ref{fig3}(a,b)
by means of two XFROG diagrams \cite{XFROG}, pertaining to $\beta
_{3}=6.98\cdot 10^{-5}$ ps$^{3}$/m and $\beta _{3}=1.75\cdot 10^{-5}$ ps$^{3}
$/m, respectively. In the former case, the chain [marked by label 2 in Fig. %
\ref{fig3}(a)] mostly belongs to the normal-GVD region, at $\lambda <$ $%
\lambda _{\mathrm{ZDP}}=790$ nm. The carrier wavelength of the ejected
soliton [marked by 1 in Fig. \ref{fig3}(a)] is shifted to $870$ nm, while
the radiation component (labeled by 3 in the figure) appears at the blue
edge of the spectrum, around $\lambda =680$ nm. In the case of the weak TOD [%
$\beta _{3}=1.75\cdot 10^{-5}$ ps$^{3}$/m, see Fig. \ref{fig3}(b)], the
chain [its segment is labeled by 2 in Fig. \ref{fig3}(b)] mostly belongs to
the anomalous-GVD region, at $\lambda >\lambda _{\mathrm{ZDP}}=761$ nm,
hence these pulses may be interpreted as being close to regular solitons.
The carrier wavelength of the ejected soliton [marked by 1 in Fig. \ref{fig3}%
(b)] is shifted to $930$ nm, while the radiation component (labeled 3 in the
figure) appears at the blue edge of the spectrum, around $\lambda =600$ nm.

\begin{figure}[tbp]
\includegraphics[scale=0.15]{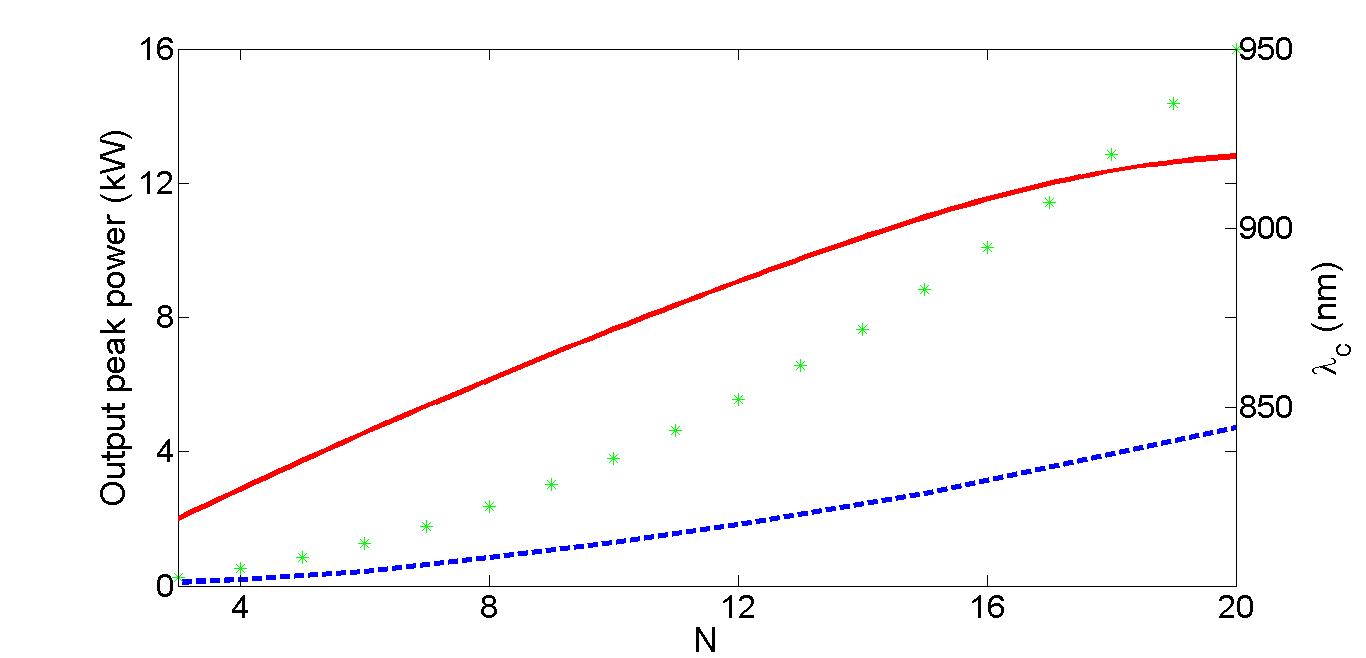}
\caption{(Color online) The wavelength shift of the tallest fundamental
soliton generated by the fission of the $N$-soliton input (\protect\ref{N}),
relative to the input wavelength of $800$ nm (the solid red curve), and the
peak power of this soliton (the dashed blue curve), vs. the input peak
power, for $\protect\beta _{3}=6.98\cdot 10^{-5}$ ps$^{3}$/m. For the sake
of comparison, the dotted green curve shows the analytical result (\protect
\ref{SY}) for the peak power of the tallest soliton in the integrable NLS
equation.}
\label{fig4}
\end{figure}

To characterize the mechanism of the TOD-driven NC formation from the
fissible $N$-solitons, we measured the peak power of the first ejected
soliton, as well as its wavelength shift, as a function of the peak power of
the originally injected $N$-soliton (\ref{N}), which is proportional to $%
N^{2}P_{0}$, while its width was fixed, as said above, to $T_{0}$ $=50$ fs.
The results are plotted in Fig. \ref{fig4} versus order $N$ of the input.

In the case of very weak TOD, the well-known analytical result of Satsuma
and Yajima \cite{Satsuma} predicts peak powers of fundamental solitons into
which the $N$th-order one is decomposed:

\begin{equation}
P^{(j)}=P_{0}(2N-2j-1)^{2},~j=0,...,N-1,  \label{SY}
\end{equation}%
where $P_{0}$ is the same as in Eq. (\ref{N}). This result, if applied to
the splitting of the $10$-soliton, shows that the largest peak power in the
set of the emerging fundamental solitons should be $3.6$ times higher than
that of the original $N$-soliton, and the energy carried away by this
tallest soliton is about $18\%$ of the total energy of the parental $N$%
-soliton. However, the relatively strong TOD makes the actual peak power of
the escaping soliton significantly lower, as seen in Fig. \ref{fig4}. For
example, the fission of the $10$-soliton displayed above produces the free
soliton with the peak power of $1$ kW, which is only $1.2$ times higher than
that of the parental $10$-soliton. On the other hand, the energy of the
single ejected soliton is about $30\%$ of initial pulse's energy. The latter
result is explained by the fact that, under the action of the strong TOD,
the collisions are not exactly elastic \cite{Energy}, allowing the ejected
soliton to collect energy donated by weaker solitons in the course of its
travel through the chain.

Due to inelastic effects of the collisions and interaction with the
conspicuous radiative component of the field, the escaping soliton
experiences additional acceleration (the wavelength shift, which is shown by
the solid red curve in Fig. \ref{fig4}). The wavelength shift starts to
saturate close to $N=20$. For even higher orders of $N$ in the input, one
can achieve still higher peak-power ratios between the tallest generated
soliton and the input, and a higher frequency shift, but, due to the
interaction between the multiple generated solitons and dispersive waves
\cite{acceleration}, the control over parameters of the generated solitons
deteriorates.

\begin{figure}[tbp]
\includegraphics[scale=0.4]{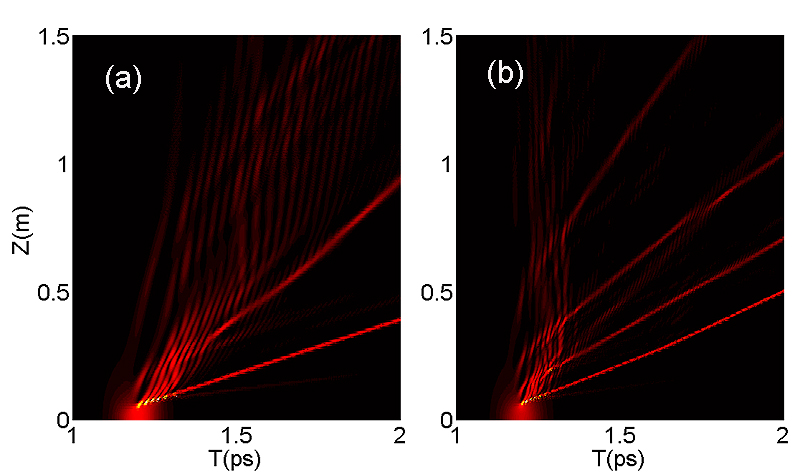}
\caption{(Color online) The fission of the $12$th-order soliton. (a) The
second fundamental soliton is ejected after the first one, at $\protect\beta %
_{3}=6.98\cdot 10^{-5}$ ps$^{3}$/m. (b) Multiple soliton ejection observed
when the TOD strength is reduced to $\protect\beta _{3}=1.74\cdot 10^{-5}$ ps%
$^{3}$/m.}
\label{fig5}
\end{figure}

At higher values of $N$ (between $10$ and $20$), the soliton-cradle effect,
in the form of a tall pulse passing the chain by means of consecutive
collisions and eventually escaping into the free space, repeats itself
several times. After the first soliton has been ejected, the cradle retains
enough strength to push additional solitons all the way through the chain,
up to the ejection accompanied by conspicuous emission of radiation, as
shown in Fig. \ref{fig5}. Reducing the TOD coefficient $\beta _{3}$, which
drives the soliton motion through the chain, makes the NC dynamical regime
less pronounced, allowing the system to increase the number of the
soliton-ejection cycles, as demonstrated in Fig. \ref{fig5}. The figure
demonstrates the fission of the $12$-soliton under the action of $\beta
_{3}=6.98\cdot 10^{-5}$ ps$^{3}$/m (a) and $\beta _{3}=1.74\cdot 10^{-5}$ ps$%
^{3}$/m (b). The effect of the multi-soliton ejection, along with the
emission of dispersive waves, from the initial $N$-soliton with high values
of $N$ manifests itself in the generation of broadband supercontinua.

To verify the robustness of the NC dynamical regimes presented above, the
simulations of Eq. (\ref{NLS}) were reproduced including other higher-order
terms, \textit{viz}., the cubic ones accounting for the self-steepening and
the Raman effect. The results demonstrate that the extra terms do not
cardinally affect the NC regime and subsequent ejection of the tallest
soliton. Actually, the higher-order nonlinear terms somewhat amplify the NC
effects, lending the ejected solitons an additional red-shift, which is
induced by the Raman term. A typical example of the fission of the $10$%
-soliton in the presence of these additional terms is displayed in Fig. \ref%
{fig6}, which demonstrates the ejection of additional solitons. It can be
compared to the $10$-soliton's fission in Fig. \ref{fig1}(d) with the same
input and the same value of the TOD coefficient, $\beta _{3}=6.98\cdot
10^{-5}$ ps$^{3}$/m, where no ejection of the second soliton was observed
for the same propagation length. Generally, the TOD-driven NC regime governs
the generation and ejection of the soliton, while its motion in the free
space is accelerated by the Raman effect.

\begin{figure}[tbp]
\includegraphics[scale=0.4]{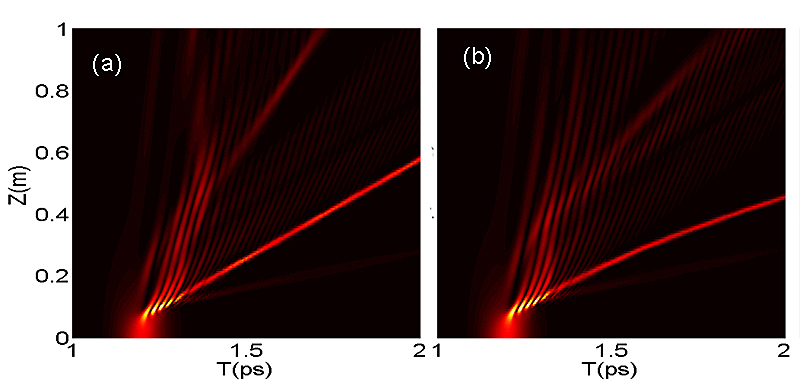}
\caption{(Color online) The fission of the $10$-soliton in the presence of
the self-steepening (a), and Raman self-frequency shift (b). In either case,
the second fundamental soliton is ejected, following the ejection of the
first soliton, after the array has passed the propagation distance $Z\simeq $
$0.5$ m.}
\label{fig6}
\end{figure}

\begin{figure}[tbp]
\includegraphics[scale=0.18]{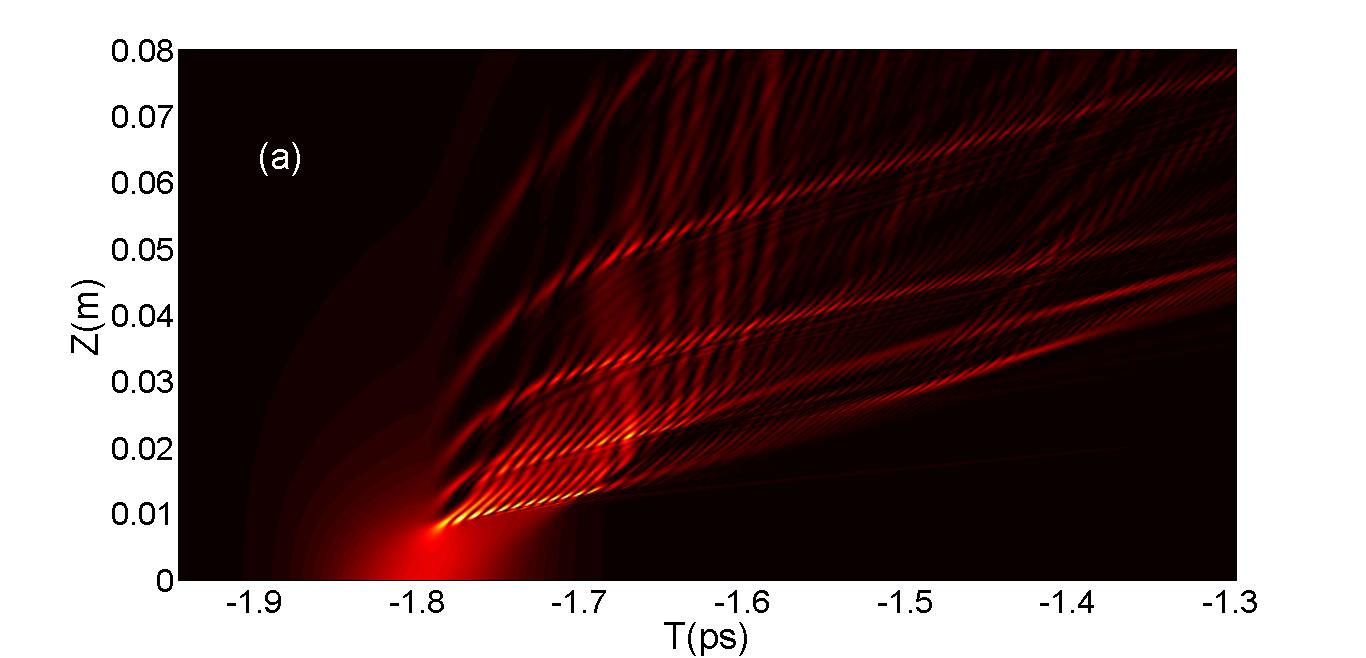} %
\includegraphics[scale=0.18]{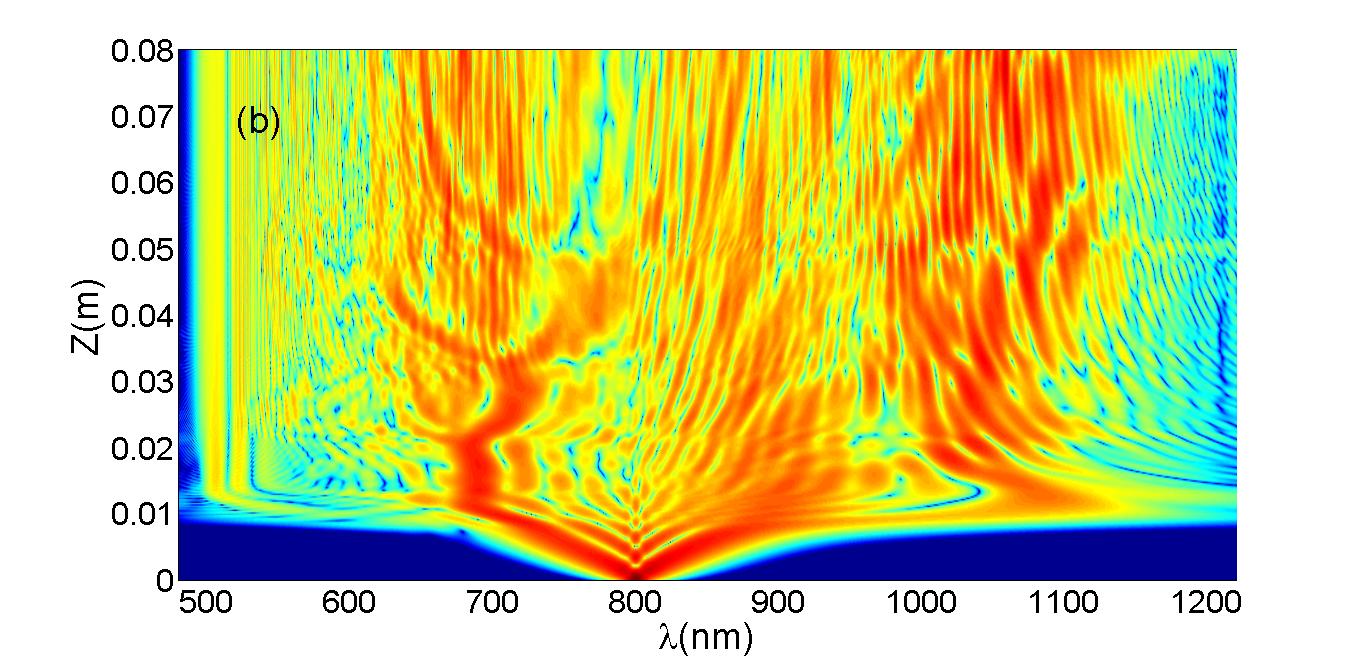}
\caption{(Color online) The fission of the $50$-soliton, as seen in the
temporal (a) and spectral (b) domains, with the subsequent formation of the
Newton's cradle and ejection of multiple solitons. The simulation was
performed in the framework of the full equation (\protect\ref{NLS}),
including the Raman and self-steepening nonlinear terms, and the
higher-order dispersive terms of up to the seventh order (see details in the
text).}
\label{fig7}
\end{figure}

It is also relevant to verify the robustness of the NC effect against the
addition of higher-order dispersive terms, which we did by simulating the
fission of the $N$-soliton in the framework of the full equation (\ref{NLS})
including the dispersive terms of up to the seventh order. The respective
dispersion coefficients were taken for the fiber model as per Ref. \cite%
{Herrmann}, at the carrier wavelength $\lambda =800$ nm (while $\lambda _{%
\mathrm{ZDP}}=790$ nm): $\beta _{2}$ $=-2.1$ fs$^{2}$/mm, $\beta _{3}$ $%
=69.83$ fs$^{3}$/mm, $\beta _{4}$ $=-73.25$ fs$^{4}$/mm, $\beta _{5}$ $%
=191.9 $ fs$^{5}$/mm, $\beta _{6}$ $=-727$ fs$^{6}$/mm, $\beta _{7}$ $=1549.4
$ fs$^{7}$/mm. The injected peak power, $26.25$ kW, corresponds to the
input's order $N=50$ (such high values of $N$ are relevant if the objective
is the generation of supercontinuum \cite{Dudleyobzor,SC}). Figures \ref%
{fig7}(a) and (b) demonstrate the corresponding fission process in the
temporal and spectral domains, respectively. In panel \ref{fig7}(a), one can
clearly observe strong signatures of the NC dynamics, in the form of narrow
collision waves ending by the ejection of several solitons along with multiple dispersive waves, thus generating
a broad supercontinuum ranging from $500$ nm to $1200$ nm.

\section{Soliton chains}

\begin{figure}[tbp]
\includegraphics[scale=0.4]{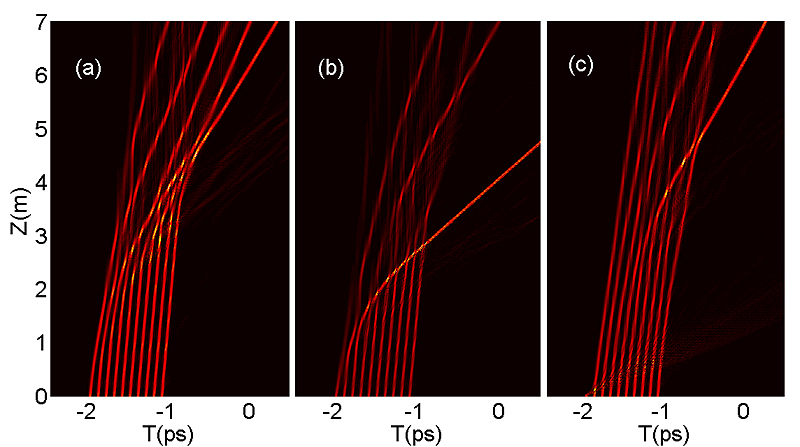}
\caption{(Color online) The dynamics of the chain of 10 identical solitons,
built as per Eq. (\protect\ref{chain}), under the action of the TOD, with $%
\protect\beta _{3}=3.5\cdot 10^{-5}$ ps$^{3}$/m, and the initial kick
applied to the leftmost soliton: (a) $\protect\omega _{k}=0$ (no kick); (b) $%
\protect\omega _{k}=-30$ ps$^{-1}$ (a weak kick); (c) $\protect\omega %
_{k}=-200$ ps$^{-1}$ (a strong kick).}
\label{fig8}
\end{figure}

\begin{figure}[tbp]
\includegraphics[scale=0.2]{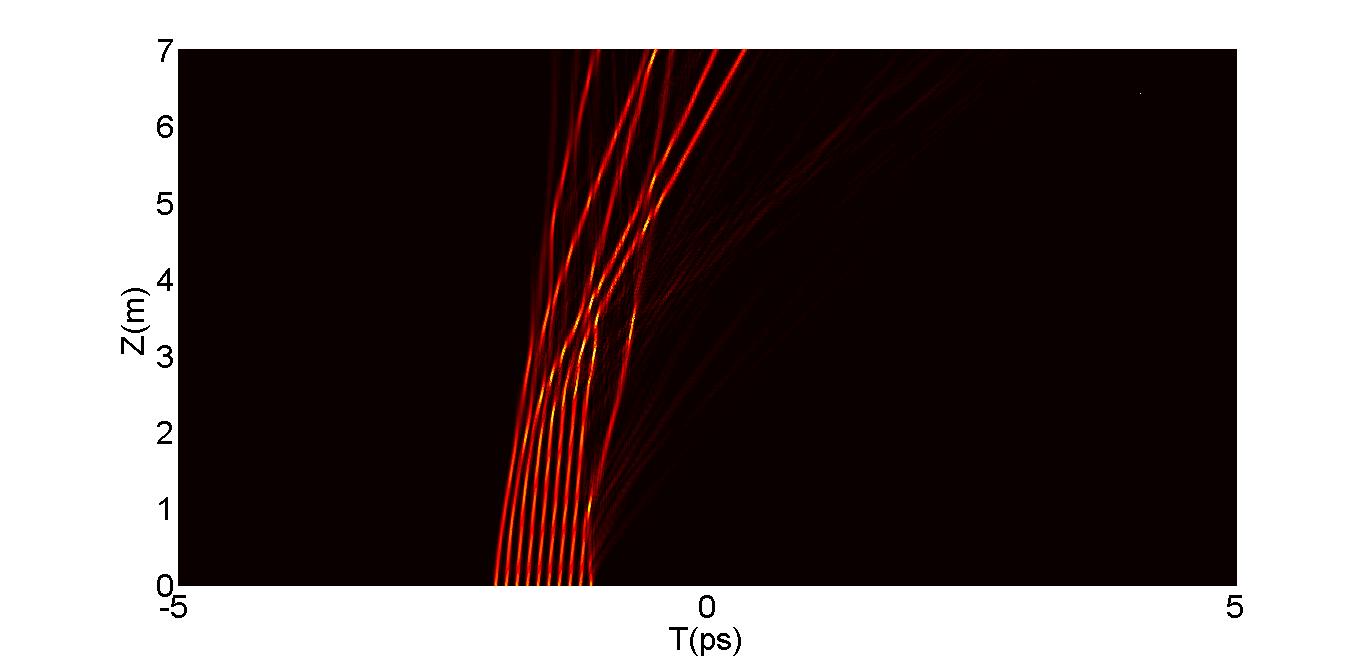}
\caption{(Color online) The same as in Fig. \protect\ref{fig9}, but when
kick $\protect\omega _{k}=+100$ ps$^{-1}$ is applied to the rightmost
soliton in the initial soliton array.}
\label{fig9}
\end{figure}

To demonstrate the generality of the realization of the TOD-driven NC
dynamics in soliton arrays, we here consider the input in the form of a
\textquotedblleft prefabricated" chain of ten identical solitary pulses with
temporal width $T_{0}=10$ fs, peak power $P_{0}=262.5$ W, and separation $%
\Delta T=5T_{0}$\textbf{\ }between them. The chain is represented by the
following input:%
\begin{equation}
u_{0}(T)=\sqrt{P_{0}}\mathrm{sech}\left( \frac{T}{T_{0}}\right) \exp
(-i\omega _{k}T)+\sqrt{P_{0}}\sum\limits_{n=2}^{10}\mathrm{sech}\left( \frac{%
T+n\Delta T}{T_{0}}\right) .  \label{chain}
\end{equation}%
Fixing the TOD coefficient here as $\beta _{3}=3.5\cdot 10^{-5}$ ps$^{3}$/m,
a negative frequency shift (\textquotedblleft kick") $\omega _{k}$ is
applied to the leftmost soliton in the chain, to initiate the NC dynamical
regime, in analogy to the classical realization of the NC in mechanics. The
result is displayed in Figs. \ref{fig8}(a-c) for the zero, weak ($\omega
_{k}=-30$ ps$^{-1}$) and strong ($\omega _{k}=-200$ ps$^{-1}$) kicks,
respectively. In addition, in Fig. \ref{fig9} another case is shown, when a
relatively strong kick ($\omega _{k}=+100$ ps$^{-1}$) is applied, in the
opposite direction, to the rightmost soliton in the chain.

The asymmetry imposed by the TOD provides energy and momentum transfer along
the chain, from left to right, through quasi-elastic repulsive collisions
between the solitons \cite{Energy}, even without any kick originally applied
to the leftmost soliton, see Fig. \ref{fig8}(a). It is relevant to mention
that the initial chain (\ref{chain}) is built of in-phase solitons. However,
the analysis of numerical data demonstrates that, as a result of the
TOD-induced asymmetry, there appear phase shifts between adjacent solitons
in the chain, and, at the evolution stage when soliton collisions take
place, the phase shifts attain values between $\pi /2$ and $3\pi /2$,
similar to the NC formed by the TOD-driven fission of the $N$-solitons, see
above.
Even in the absence of the initial kick, Fig. \ref{fig8}(a) demonstrates a
dynamical regime which resembles the NC dynamics in the soliton chains which
was observed above, namely, the propagation of a narrow collision wave
through the chain, that ends up by the ejection of a fast soliton. The
characteristic features of the NC dynamics become more prominent under the
action of the initial kick, see Fig. \ref{fig8}(b). The situation is,
however, different when the first soliton is set in motion by a strong kick.
The initial violent collision leads to rapid destruction of the incident
soliton, see Fig. \ref{fig8}(c), and the chain subsequently demonstrates a
less pronounced quasi-NC regime, more similar to that in the absence of the
kick, cf. Fig. \ref{fig8}(a). Thus, dealing with the chain of identical
solitons, we conclude that there exists a range of values of the initial
kick which gives rise to the most well-pronounced NC behavior. On the other
hand, it is relevant to mention that taking $\beta _{3}$ twice as large as
in these simulations produces a picture which, without any kick, is quite
similar to the one generated by the applied kick in Fig. \ref{fig8}(b).

Kicking the rightmost soliton in the opposite direction leads to its
collision with the neighbor and their merger into a single soliton. After
that, the NC develops in essentially the same form as in the absence of the
kick, see Fig. \ref{fig9}.

\begin{figure}[tbp]
\includegraphics[scale=0.2]{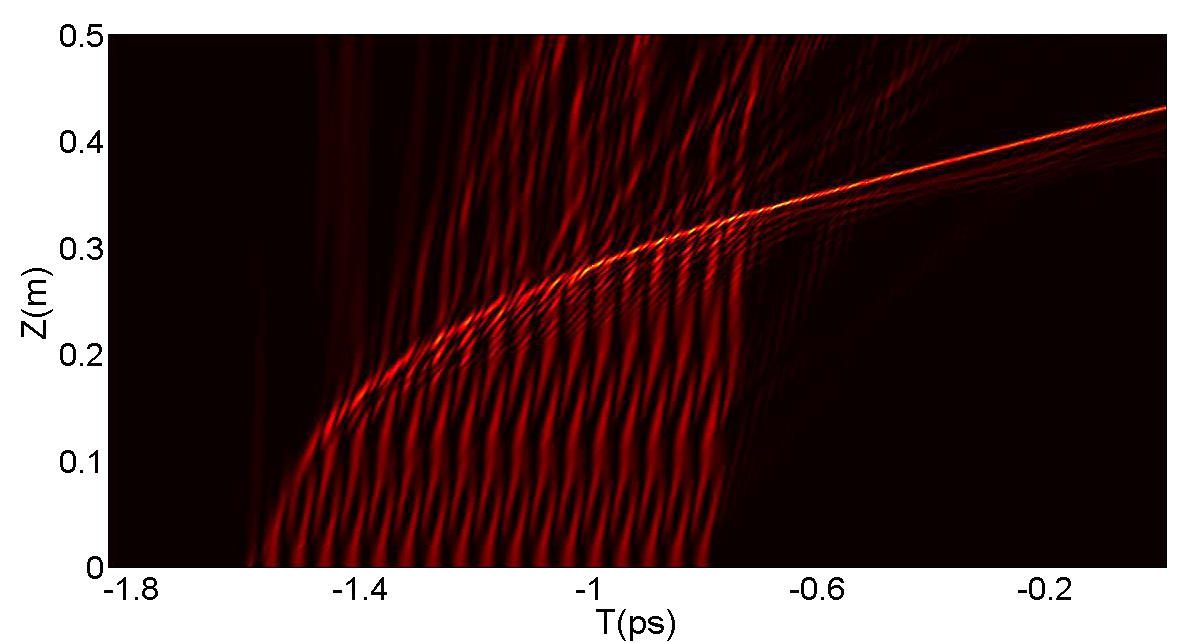}
\caption{(Color online) The NC emulated by launching as an input a
cosinusoidal wave. The truncation of the input was done at the time interval
of about 0.8 ps. This simulation was performed with TOD coefficient $\protect%
\beta _{3}=6.98\cdot 10^{-5}$ ps$^{3}$/m.}
\label{fig10}
\end{figure}

For emulating the NC structure, the initial periodic pattern does not need
to be built as a chain of solitons. As Fig. \ref{fig10} shows, one can use
as an input simply a truncated cosinusoidal wave, $u_{0}(T)=\sqrt{P_{0}}\cos
(T/\tilde{T}_{0})$. The NC dynamics, resulting in the final release of the
accelerated soliton with an enhanced power, is well observed in this case
too.

Lastly, it is relevant to mention that a recently investigated regime of the
evolution of the input in the form of a truncated Airy function, $%
u_{0}(T)=A_{0}\mathrm{Ai}(T)\exp (aT)$, in the framework of the integrable
NLS\ equation (without the TOD term), which gives rise to ejection of
individual solitons from the Airy pattern \cite{Marom}, also demonstrates
features similar to those expected from the NC dynamics in a multi-pulse
chain. Another relevant example is provided by recent works which simulated
the splitting of a broad input into soliton arrays in models with the
quadratic (second-harmonic-generating nonlinearity) \cite{Bache}: in some
cases, they display the release of Cherenkov wave packets, which resemble
the ejection of the free soliton form the NC soliton array, which was
reported above.

\section{Conclusion}

We have introduced the optical analog of the NC (Newton's cradle) in the
form of soliton arrays, or chains of optical pulses. The characteristic
feature of the NC, i.e., the transmission of a narrow wave of collisions,
which is realized as the passage of the tallest soliton through the whole
array, from left to right, is driven by the TOD, and ends by release of the
tallest soliton, with a conspicuous frequency shift. Due to inelasticity of
soliton collisions under the action of TOD, the passing soliton collects
energy in the course of its motion through the array. We have demonstrated
that the NC, built of optical pulses with different amplitudes, is naturally
generated by the TOD-driven fission of $N$-solitons. Similar NC dynamical
regimes are induced as well by the TOD acting on arrays of identical
solitons. For $N$ large enough, the NC may support the multiple passage and
eventual ejection of solitons with enhanced power, through their consecutive collisions
with other pulses that stay bound in the long chain. Along with multiple dispersive waves emitted the ejected solitons generate a broadband supercontinuum.
The NC-forming mechanism is robust against the inclusion of the Raman and self-steepening effects, as well as dispersion
terms of order higher than three.

While the similarity of the dynamical regimes reported above to the
classical mechanical NC is far from being exact, the features demonstrated
by the arrays of solitons, suggest that the original NC concept may be
extended for nonlinear-wave settings, acquiring new properties: the soliton
arrays building the NC as a result of the fission of $N$-soliton input are
naturally non-uniform, and the tallest soliton, moving across the array,
effectively passes through other solitons, which is possible for
quasi-particles, but impossible for hard beads in the mechanical NC.
Furthermore, the passing soliton collects the energy and momentum from other
solitons, and is eventually released along with conspicuous amounts of
radiation, which lends the NC dynamics in soliton arrays novel inelastic
features. The dynamical regimes that proceed via the formation of the
soliton NCs may find important applications in optics, such as optimization
of the supercontinuum generation.

It is expected that soliton-based NCs may also be constructed in other
models of nonlinear optics, such as chains of gap solitons in Bragg gratings
or photonic crystals. In particular, the breakup of higher-order spatial
solitons under the action of nonlinear absorption, demonstrated in Ref. \cite%
{Silberberg}, may be related to this phenomenology.

\section{Acknowledgements}

We appreciate valuable discussions with Anatoly Efimov. The work of AVY was supported by the FCT grant PTDC/FIS/112624/2009 and PEst-OE/FIS/UI0618/2011.

\end{document}